\documentclass[pdftex, pra, twocolumn, floatfix, superscriptaddress, nofootinbib]{revtex4}

\usepackage{times}
\usepackage{amsfonts}
\usepackage{amssymb}
\usepackage{amsmath}
\usepackage{graphicx}
\usepackage{subfigure}
\usepackage[colorlinks=true, breaklinks=true, linkcolor=blue, citecolor=blue, urlcolor=blue]{hyperref}
\usepackage{bbm}

\newcommand{\doi}[2]{\href{https://doi.org/#1}{#2}}
\newcommand{\arxiv}[1]{\href{https://arxiv.org/abs/#1}{arXiv:#1}}

\begin{document}

\title{Spin-chain model for strongly interacting one-dimensional Bose-Fermi mixtures}

\author{F. Deuretzbacher}
\email{frank.deuretzbacher@itp.uni-hannover.de}
\affiliation{Institut f\"ur Theoretische Physik, Leibniz Universit\"at Hannover, Appelstrasse 2, DE-30167 Hannover, Germany}

\author{D. Becker}
\affiliation{I. Institut f\"ur Theoretische Physik, Universit\"at Hamburg, Jungiusstrasse 9, DE-20355 Hamburg, Germany}

\author{J. Bjerlin}
\affiliation{Mathematical Physics and NanoLund, LTH, Lund University, SE-22100 Lund, Sweden}

\author{S. M. Reimann}
\affiliation{Mathematical Physics and NanoLund, LTH, Lund University, SE-22100 Lund, Sweden}

\author{L. Santos}
\affiliation{Institut f\"ur Theoretische Physik, Leibniz Universit\"at Hannover, Appelstrasse 2, DE-30167 Hannover, Germany}

\begin{abstract}
Strongly interacting one-dimensional (1D) Bose-Fermi mixtures form a tunable XXZ spin chain. Within the spin-chain model developed here, all properties of these systems can be calculated from states representing the ordering of the bosons and fermions within the atom chain and from the ground-state wave function of spinless noninteracting fermions. We validate the model by means of an exact diagonalization of the full few-body Hamiltonian in the strongly interacting regime. Using the model, we explore the phase diagram of the atom chain as a function of the boson-boson (BB) and boson-fermion (BF) interaction strengths and calculate the densities, momentum distributions, and trap-level occupancies for up to 17 particles. In particular, we find antiferromagnetic (AFM) and ferromagnetic (FM) order and a demixing of the bosons and fermions in certain interaction regimes. We find, however, no demixing for equally strong BB and BF interactions in agreement with earlier calculations that combined the Bethe ansatz with a local-density approximation.
\end{abstract}

\maketitle

\section{Introduction}

Ultracold atoms are ideally suited to study strongly correlated one-dimensional (1D) systems due to their high degree of control and tunability \cite{Cazalilla11, Guan13}. These advantageous features have led to the observation of the Tonks-Girardeau gas \cite{Kinoshita04, Paredes04}, the controlled preparation of a highly excited super-Tonks gas \cite{Haller09, Astrakharchik05}, undamped dynamics in strongly interacting 1D Bose gases \cite{Kinoshita06}, and the deterministic preparation of 1D few-fermion systems with tunable interactions \cite{Serwane11, Zuern12, Zuern13, Wenz13, Murmann15}. Moreover, it became possible to realize a variety of artificial 1D systems consisting, e.g., of atoms with a large spin \cite{Pagano14} or Bose-Fermi mixtures with mixed statistics \cite{Schreck01}.

These developments have renewed the interest in Girardeau's Bose-Fermi mapping for 1D spinless bosons with infinite $\delta$ repulsion \cite{Girardeau60} leading to generalizations for Bose-Fermi mixtures \cite{Girardeau07}, spin-1 bosons \cite{Deuretzbacher08}, and spin-1/2 fermions \cite{Guan09}. Only recently it was found that these exact solutions are also useful for the perturbative treatment of strongly interacting 1D systems \cite{Volosniev14}. Different from one-component systems, the ground state of multicomponent systems with infinite $\delta$ repulsion is highly degenerate \cite{Girardeau07, Deuretzbacher08, Guan09}. This is due to the fact that strongly interacting 1D particles localize and arrange themselves in a spin chain \cite{Deuretzbacher08, Matveev08, Deuretzbacher14}. This offers the exciting possibility to study quantum magnetism without the need for an optical lattice \cite{Deuretzbacher14, Volosniev15, Yang15, Massignan15, Yang16a}.

Theoretical studies of Bose-Fermi mixtures in optical lattices predicted composite fermions consisting of one fermion and one ore more bosons, or, respectively, bosonic holes \cite{Lewenstein04} and polarons \cite{Mathey07}. In addition, pairing, collapse, and demixing can occur in homogeneous 1D systems of strongly interacting bosons and fermions \cite{Cazalilla03}. For equally strong boson-boson (BB) and boson-fermion (BF) $\delta$ interactions, the model can be solved exactly via the Bethe ansatz \cite{Imambekov06, Guan08}. Selected states of the degenerate ground-state multiplet have been constructed in the Tonks-Girardeau regime of infinite $\delta$ repulsion \cite{Girardeau07} and classified using Young's tableaux \cite{Fang11}. Recently, all states of the multiplet have been constructed for few (up to 6) particles \cite{Hu16a, Dehkharghani17, Zinner15} and strongly interacting mixtures with additional weak $p$-wave interactions have been studied \cite{Hu16b, Yang16b}.

Here, we develop a spin-chain model for 1D Bose-Fermi mixtures with nearly infinite $\delta$ interactions \cite{AncillaryFiles}. We check the validity of the model by diagonalizing the full few-body Hamiltonian numerically in the strongly interacting regime. Using the spin-chain model, we then calculate the ground-state densities, momentum distributions, and occupancies of the harmonic-trap levels for atom chains consisting of up to 17 particles. Moreover, we determine the ground-state phases of these atom chains, finding antiferromagnetic (AFM) and ferromagnetic (FM) order and a demixing of the bosons and fermions for particular values of the BB and BF interaction strengths. However, no demixing is found for equally strong BB and BF interactions although the bosons are predominantly in the trap center and the fermions are predominantly at the edges of the trap \cite{Imambekov06}.

\section{Spin-chain model}

We consider a 1D mixture of $N_B$ bosons and $N_F$ fermions (total particle number $N=N_B+N_F$). Both species are assumed to have the same masses, $M_B = M_F = M$, and experience the same trapping potential $V(z)$. The bosons interact with each other through a $\delta$ potential of strength $g_{BB}$ and with the fermions through a $\delta$ potential of strength $g_{BF}$. The many-body Hamiltonian of the system is given by
\begin{eqnarray} \label{eq-HBF}
H & \!\! = & \!\! \sum_{i=1}^N \left[ -\frac{\hbar^2}{2M} \frac{\partial^2}{\partial z_i^2} + V(z_i) \right] \! + g_{BB} \! \sum_{i=1}^{N_B - 1} \! \sum_{j=i+1}^{N_B} \! \delta (z_i - z_j) \nonumber \\
& & \!\! + \; g_{BF} \sum_{i=1}^{N_B} \sum_{j = N_B + 1}^N \delta (z_i - z_j) .
\end{eqnarray}
The interaction strengths~$g_{BB}$ and $g_{BF}$ are freely tunable through a magnetic Feshbach resonance \cite{Inouye98} and through the strong radial confinement \cite{Olshanii98}.

Here, we focus on the strongly interacting regime, where the absolute value of both interaction strengths is large, i.~e., $|g_{BB}| \approx \infty$ and $|g_{BF}| \approx \infty$. Furthermore, we consider only the highly excited super-Tonks states \cite{Haller09, Astrakharchik05, Zuern12} if at least one of the interaction strengths is attractive.\footnote{Super-Tonks states may be prepared by ramping adiabatically across a confinement-induced resonance \cite{Haller09, Zuern12, Comment1}.} Under these conditions, the atoms order in a row and form a spin chain \cite{Deuretzbacher08, Matveev08, Deuretzbacher14, Comment1}. An arbitrary state of the spin chain is given by
\begin{equation}
| \chi \rangle = \sum_{m_1, \dotsc, m_N} c_{m_1, \dotsc, m_N} | m_1, \dotsc, m_N \rangle ,
\end{equation}
where each basis state $| m_1, \dotsc, m_N \rangle$ with $m_i = B, F$ corresponds to a particular ordering of the bosons ($B$) and fermions ($F$) and can be constructed from the wave function of $N$ spinless noninteracting fermions~\cite{Fang11} (see Appendix~\ref{app-sector-wf}).

Nearest neighboring particles of the spin chain interact with each other through the effective Hamiltonian \cite{Hu16b}
\begin{eqnarray} \label{eq-Heff}
\!\!\! H_\mathrm{eff} & \! = & \! E_F \openone - 2 \sum_{i=1}^{N-1} J_i^{(BB)} |B\rangle_i |B\rangle_{i+1} \langle B|_i \langle B|_{i+1} \nonumber \\
& & \! - \sum_{i=1}^{N-1} J_i^{(BF)} \Bigl( |B\rangle_i |F\rangle_{i+1} + |F\rangle_i |B\rangle_{i+1} \Bigr) \nonumber \\
& & \! \times \Bigl( \langle B|_i \langle F|_{i+1} + \langle F|_i \langle B|_{i+1} \Bigr) ,
\end{eqnarray}
as shown in Appendix~\ref{app-Heff}. Here, $E_F$ is the ground-state energy of $N$ spinless noninteracting fermions in the trapping potential $V(z)$ and $J_i^{(BB)}$ and $J_i^{(BF)}$ are the exchange coefficients of nearest neighbor bosons or bosons and fermions, respectively. The exchange coefficients are given by $J_i^{(BB)} = C_i / g_{BB}$ and $J_i^{(BF)} = C_i / g_{BF}$ with \cite{Volosniev14, Deuretzbacher14}
\begin{equation}
C_i = \frac{N! \hbar^4}{M^2} \! \int \! dz_1 \dotsi dz_N \delta(z_i-z_{i+1}) \theta (z_1,\dotsc,z_N) \! \left| \frac{\partial \psi_F}{\partial z_i} \right|^2 \!\! ,
\end{equation}
where $\theta (z_1, \dotsc , z_N) = 1$ if $z_1 < \dotsb < z_N$, and zero otherwise, and where $\psi_F$ is the ground-state wave function of $N$ spinless noninteracting fermions in the trap $V(z)$. The $C_i$ can be efficiently calculated for large $N$ \cite{Deuretzbacher16, Loft16, Yang16c}.

By identifying bosons and fermions with pseudospin-up and -down particles, respectively, Eq.~(\ref{eq-Heff}) can be rewritten in terms of the Pauli matrices $\sigma_x^{(i)}$, $\sigma_y^{(i)}$, and $\sigma_z^{(i)}$:
\begin{eqnarray} \label{eq-XXZ}
& & \mspace{-20mu} H_\mathrm{eff} = - \frac{1}{2} \sum_{i=1}^{N-1} \biggl\{ J_i^{(BF)} \left[ \sigma_x^{(i)} \sigma_x^{(i+1)} + \sigma_y^{(i)} \sigma_y^{(i+1)} \right] \nonumber \\
& & \mspace{-20mu} + \left[ J_i^{(BB)} - J_i^{(BF)} \right] \sigma_z^{(i)} \sigma_z^{(i+1)} + J_i^{(BB)} \left[ \sigma_z^{(i)} + \sigma_z^{(i+1)} \right] \biggr\} . \nonumber \\
& &
\end{eqnarray}
Here, we neglected the diagonal matrix
\begin{equation}
\left\{ E_F - \frac{1}{2} \sum_{i=1}^{N-1} \left[ J_i^{(BB)} + J_i^{(BF)} \right] \right\} \openone .
\end{equation}
Equation~(\ref{eq-XXZ}) is the Hamiltonian of an XXZ spin chain in an inhomogeneous magnetic field along the $z$ axis. Similar effective Hamiltonians have been derived for strongly interacting Bose-Bose mixtures \cite{Volosniev15, Massignan15, Yang16a} and strongly interacting mixtures with weak $p$-wave interactions \cite{Hu16b, Yang16b}.

The densities of the bosons (${m=B}$) and fermions (${m=F}$) are given by \cite{Deuretzbacher08}
\begin{equation}
\rho_m (z) = \sum_{i=1}^N \rho^{(i)}(z) \rho_m^{(i)}
\end{equation}
with the probability to find the $i$th particle at position~$z$,
\begin{equation}
\rho^{(i)}(z) = N! \int d z_1 \dotsi d z_N \delta (z-z_i) \theta (z_1, \dotsc , z_N) |\psi_F|^2 ,
\end{equation}
and the probability that the $i$th particle is a boson ($m=B$) or fermion ($m=F$),
\begin{equation}
\rho_m^{(i)} = \! \sum_{m_1, \dotsc, m_N} \bigl| \langle m_1, \dotsc, m_N | \chi \rangle \bigr|^2 \delta_{m, m_i} .
\end{equation}
The one-body density matrix of the bosons ($m=B$) and fermions ($m=F$) is given by
\begin{equation} \label{eq-one-body}
\rho_m (z,z') = \sum_{i,j=1}^N \rho^{(i,j)} (z,z') \rho_m^{(i,j)}
\end{equation}
with \cite{Yang15, Deuretzbacher16}
\begin{eqnarray}
& & \mspace{-20mu} \rho^{(i,j)}(z,z') = N! \int dz_1 \dotsi dz_{i-1} dz_{i+1} \dotsi dz_N \nonumber \\
& & \times \Bigl[ \theta (z_1, \dotsc , z_N) \bigl| \psi_F (z_1, \dotsc , z_N) \bigr| \Bigr]_{z_i=z} \nonumber \\
& & \times \Bigl[ \theta (z_{P_{i, \dotsc, j}(1)}, \dotsc , z_{P_{i, \dotsc, j}(N)}) \bigl| \psi_F (z_1, \dotsc , z_N) \bigr| \Bigr]_{z_i=z'} \nonumber \\
& &
\end{eqnarray}
and
\begin{equation} \label{eq-rhoij}
\rho^{(i,j)}_m = \langle\chi|m\rangle_i \langle m|_i \hat P_{i,\dotsc,j}^{(BF)} |\chi\rangle ,
\end{equation}
as shown in Appendix~\ref{app-one-body}. Here, we defined $\hat P_{i,\dotsc,j}^{(BF)} = (-1)^{N_\mathrm{tr}} \hat P_{i,\dotsc,j}$ with the loop permutation operator $\hat P_{i,\dotsc,j}$, which moves a particle from position~$j$ to position~$i$ (see Appendix~\ref{app-sector-wf} for details). $N_\mathrm{tr}$ is the number of transpositions of neighboring fermions when $\hat P_{i,\dotsc,j}$ acts on $| m_1, \dotsc, m_N \rangle$. The momentum distributions and occupancies of the trap levels are related to the one-body density matrices by
\begin{equation}
\rho_m(k) = \frac{1}{2\pi} \int dz dz' e^{\mathbbm{i} k (z-z')} \rho_m (z,z')
\end{equation}
and
\begin{equation}
\rho_m(n) = \int dz dz' \phi_n(z) \phi_n^*(z') \rho_m(z,z')
\end{equation}
with the eigenfunctions $\phi_n(z)$ of the trap $V(z)$.

\begin{figure}
\begin{center}
\includegraphics[width = \columnwidth]{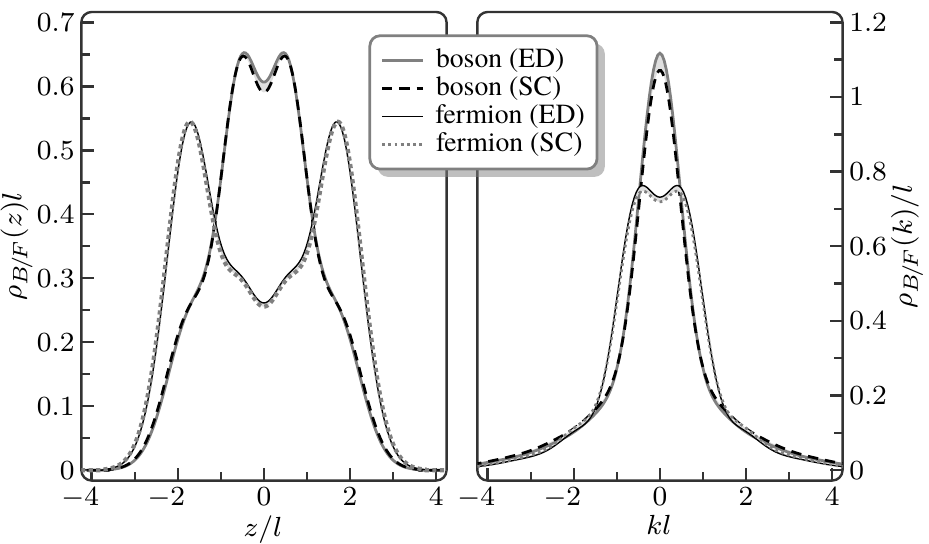}
\caption{Comparison of the ground-state densities (left) and momentum distributions (right) of a mixture of two bosons and two fermions (2B2F mixture) in a harmonic trap calculated by means of an exact diagonalization (ED) of the full few-body Hamiltonian (solid lines) and from the spin-chain (SC) model (dashed lines) for equal BB and BF interaction strengths ($g_{BB} = g_{BF} = 15 \hbar \omega l$). $\omega$ and $l$ are the frequency and length scale of the harmonic oscillator.}
\label{fig-comparison}
\end{center}
\end{figure}

We have tested the validity of the spin-chain model by comparing its results to those of an exact diagonalization of the full few-body Hamiltonian~(\ref{eq-HBF}) for up to four particles in a harmonic trap. Both approaches should lead to the same results in the Tonks-Girardeau regime, $g_{BB} = g_{BF} > 10 \hbar \omega l$ \cite{Deuretzbacher07, Deuretzbacher14} $\bigl[ \omega$ is the frequency and $l = \sqrt{\hbar / (m \omega)}$ is the length scale of the harmonic oscillator$\bigr]$. Indeed, the comparison showed excellent agreement for the spectrum, the densities, and the momentum distributions for mixtures consisting of one boson and three fermions (1B3F mixture), two bosons and two fermions (2B2F mixture), and three bosons and one fermion (3B1F mixture). As an example, we show in Fig.~\ref{fig-comparison} the result of such a comparison for the ground-state densities and momentum distributions of a 2B2F mixture for equally strong BB and BF interactions, $g_{BB} = g_{BF} = 15 \hbar \omega l$. These ground-state densities agree with Refs. \cite{Hu16a, Dehkharghani17}.

\begin{figure}[h]
\begin{center}
\includegraphics[width = 0.8 \columnwidth]{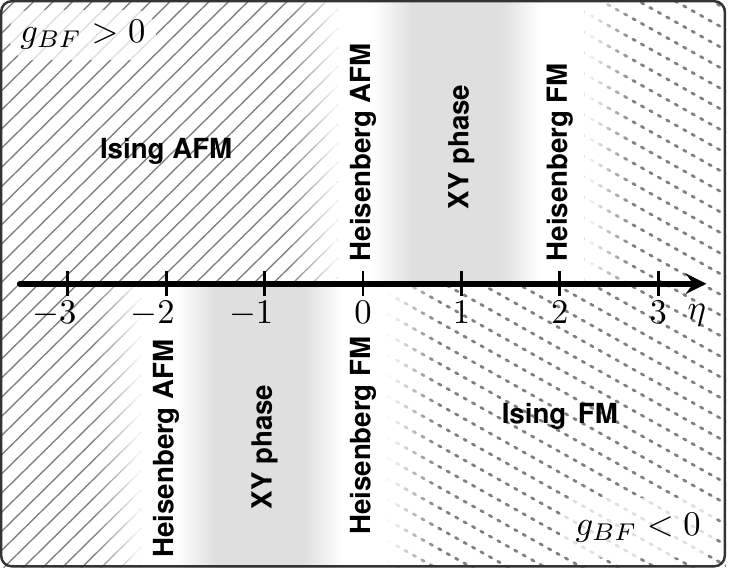}
\caption{Phase diagram of the Bose-Fermi chain as a function of $\eta = |g_{BF}| / g_{BB}$ for $g_{BF}>0$ (upper part) and $g_{BF}<0$ (lower part).}
\label{fig-phases}
\end{center}
\end{figure}

\section{Phases and densities}
\label{sec-phases}

The ground-state phases of the effective Hamiltonian~(\ref{eq-Heff}) or, respectively, (\ref{eq-XXZ}) are determined by the interplay of the BB and BF interactions. In particular, we distinguish five different phases, as shown in the phase diagram in Fig.~\ref{fig-phases}, which follow from the phases of the homogeneous XXZ chain in the absence of an external magnetic field \cite{Mikeska04}.\footnote{A homogeneous external magnetic field has no effect due to the conserved total magnetization of the system in the $z$ direction.} For dominant BB exchange couplings, which corresponds to the parameter regime $|\eta| = |J_i^{(BB)} / J_i^{(BF)}| \gg 1$, the spin chain is in the Ising AFM ($J_i^{(BB)} < 0$) or FM ($J_i^{(BB)} > 0$) state. For $J_i^{(BB)} \approx J_i^{(BF)}$ (gray shaded regions around $\eta = \pm 1$), the spin chain is in the XY phase. These phases are characterized by strong FM ($J_i^{(BF)} > 0$) or AFM ($J_i^{(BF)} < 0$) $xy$ correlations. At the edges of the XY phases ($\eta = 0, \pm 2$) the system is in the Heisenberg AFM or FM phases. Note that we consider the highly excited (metastable) super-Tonks states \cite{Haller09, Astrakharchik05} in the regime of attractive interactions and, therefore, do not obtain collapse and pairing \cite{Cazalilla03}.

\begin{figure}
\begin{center}
\includegraphics[width = \columnwidth]{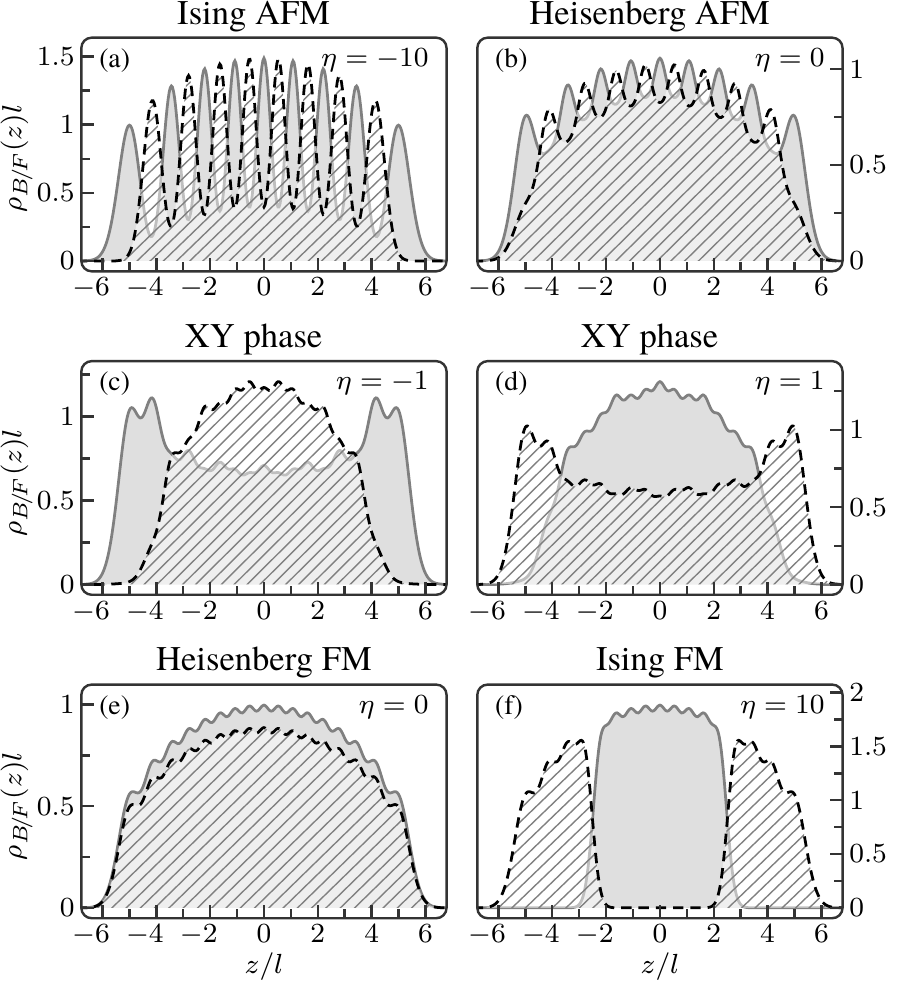}
\caption{Ground-state densities of a harmonically trapped Bose-Fermi mixture consisting of nine bosons (solid) and eight fermions (dashed) for $g_{BF}<0$ (left) and $g_{BF}>0$ (right). The different phases are described in the text. $l$ is the harmonic oscillator length.}
\label{fig-density}
\end{center}
\end{figure}

{\it Ising FM and AFM order $(|\eta| \gg 1)$.} Let us first consider the case $J_i^{(BF)} = 0$, $J_i^{(BB)} \neq 0$, in which the effective Hamiltonian~(\ref{eq-Heff}) is diagonal. In that case, for $J_i^{(BB)} > 0$, the energy is minimized if all bosons are next to each other and in the trap center (largest $C_i$). A typical ground state is therefore of the form $|F,F,B,B,B,B,F,F\rangle$, i.~e., the bosons are separated from the fermions, the bosons are in the trap center, and the fermions are at the edges of the trap. This separation of the bosons from the fermions is clearly visible in the densities of Fig.~\ref{fig-density}(f). In the opposite case of negative BB exchange coefficients, $J_i^{(BB)} < 0$, the energy is minimized if the bosons are not next to each other and hence a typical ground state is of the form $|B,F,B,F,B,F,B\rangle$, as in the ground state of an Ising AFM chain. In that regime, the densities look, therefore, like those of Fig.~\ref{fig-density}(a). The same or similar limiting phases have been found in related mixtures \cite{Massignan15, Zinner15, Hu16b, Yang16b}.

{\it Heisenberg AFM and FM order $(\eta = 0)$.} Let us now consider the case $J_i^{(BB)} = 0$, $J_i^{(BF)} \neq 0$, for which the spin-chain Hamiltonian~(\ref{eq-XXZ}) takes the form
\begin{equation}
H_\mathrm{eff} = - \! \sum_{i=1}^{N-1} \! \frac{J_i^{(BF)}}{2} \! \left[ \sigma_x^{(i)} \sigma_x^{(i+1)} + \sigma_y^{(i)} \sigma_y^{(i+1)} - \sigma_z^{(i)} \sigma_z^{(i+1)} \right] \! .
\end{equation}
By performing the unitary transformation
\begin{equation} \label{eq-transform}
\sigma_x^{(i)} \rightarrow (-1)^i \sigma_x^{(i)} \quad \sigma_y^{(i)} \rightarrow (-1)^i \sigma_y^{(i)} \quad \sigma_z^{(i)} \rightarrow \sigma_z^{(i)}
\end{equation}
we obtain
\begin{equation} \label{eq-Heisenberg}
\widetilde H_\mathrm{eff} = \sum_{i=1}^{N-1} \frac{J_i^{(BF)}}{2} \vec \sigma^{(i)} \cdot \vec \sigma^{(i+1)} ,
\end{equation}
which is the Heisenberg Hamiltonian. Therefore, the ground state is AFM for $J_i^{(BF)} > 0$ and FM for $J_i^{(BF)} < 0$. Typical densities are shown in Figs.~\ref{fig-density}(b) and (e). The spin-spin correlations $\sigma_z^{(i)} \sigma_z^{(j)}$ alternate in sign, $\propto (-1)^{i-j}$, and decay with distance $|i-j|$ in the AFM state, while staying constant in the FM state. However, because of the transform~(\ref{eq-transform}), the spin-spin correlations in the $xy$ plane do not alternate in sign in the AFM state, but instead, they alternate in the FM state.

{\it XY phases $(\eta = \pm 1)$.} Let us finally discuss the cases $J_i^{(BB)} = J_i^{(BF)}$ ($g_{BB} = g_{BF}$). The repulsive case, $g_{BB} = g_{BF} > 0$, is exactly solvable for any value of the interaction strength if $V(z)=0$ \cite{Imambekov06, Guan08}. Combining the exact solution of the homogeneous system with a local density approximation, one finds that the bosons and fermions do not demix, but the bosons are predominantly in the trap center and the fermions are predominantly at the edges of the trap. We find the same result and the density profiles are in excellent agreement with Ref. \cite{Imambekov06}, see Fig.~\ref{fig-density}(d). For attractive interactions, $g_{BB} = g_{BF} < 0$, the situation is reversed with the fermions (bosons) sitting predominantly at the center (edges) of the harmonic trap, see Fig.~\ref{fig-density}(c). We note that the bosonic (fermionic) density of the $\eta=1$ case would exactly equal the fermionic (bosonic) density of the $\eta=-1$ case, if the particle numbers would be equal, i.e., $N_B=N_F$.

This symmetry can be understood as follows: For $J_i^{(BB)} = J_i^{(BF)} > 0$, Eq.~(\ref{eq-XXZ}) takes the form of an XX Hamiltonian with an inhomogeneous effective magnetic field pointing along the $+z$ direction,
\begin{eqnarray}
H_\mathrm{eff} & = & - \sum_{i=1}^{N-1} \frac{|J_i^{(BF)}|}{2} \left[ \sigma_x^{(i)} \sigma_x^{(i+1)} + \sigma_y^{(i)} \sigma_y^{(i+1)} \right] \nonumber \\
& & - \sum_{i=1}^{N-1} \frac{|J_i^{(BB)}|}{2} \left[ \sigma_z^{(i)} + \sigma_z^{(i+1)} \right] ,
\end{eqnarray}
while for $J_i^{(BB)} = J_i^{(BF)} < 0$, after performing the transformation~(\ref{eq-transform}), the field points into the $-z$ direction,
\begin{eqnarray}
\widetilde H_\mathrm{eff} & = & - \sum_{i=1}^{N-1} \frac{|J_i^{(BF)}|}{2} \left[ \sigma_x^{(i)} \sigma_x^{(i+1)} + \sigma_y^{(i)} \sigma_y^{(i+1)} \right] \nonumber \\
& & + \sum_{i=1}^{N-1} \frac{|J_i^{(BB)}|}{2} \left[ \sigma_z^{(i)} + \sigma_z^{(i+1)} \right] .
\end{eqnarray}
In the first case, the bosons (pseudospin-up) are moved to the trap center, since the $|J_i^{(BB)}|$ are largest there, while in the latter case, the fermions (pseudospin-down) are moved to the trap center. Moreover, both Hamiltonians can be transformed into each other by exchanging the bosons with the fermions, which explains the symmetry of the density distributions.

\begin{figure}
\begin{center}
\includegraphics[width = \columnwidth]{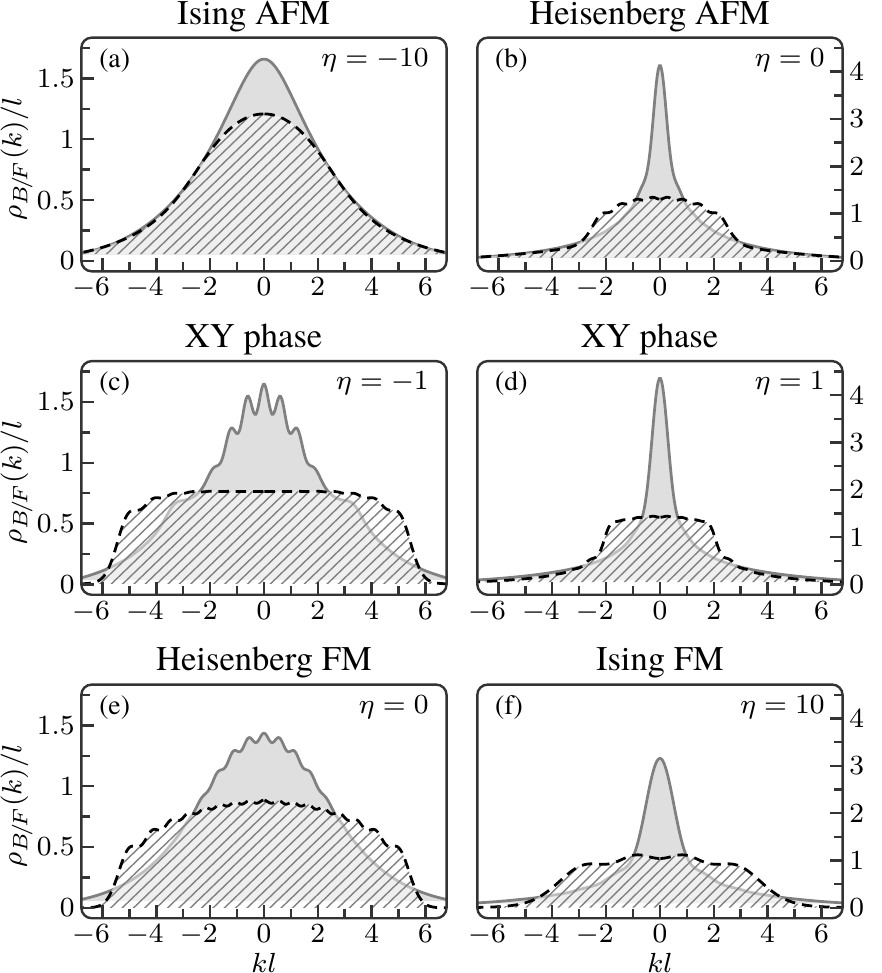}
\caption{Momentum distributions of nine bosons (solid) and eight fermions (dashed) in a harmonic trap for $g_{BF}<0$ (left column) and $g_{BF}>0$ (right column). See the text for a description of the different phases. $l$ is the harmonic oscillator length.}
\label{fig-momentum}
\end{center}
\end{figure}

\section{Momentum distributions and occupancies}

The momentum distributions and the occupancies of the harmonic-trap levels are important observables, which can be measured in the experiment \cite{Pagano14, Murmann15, Meinert16}. These distributions depend strongly on the degree of exchange symmetry of the spatial many-body wave function and can also be used as a probe for the magnetic structure of the spin chain \cite{Murmann15, Deuretzbacher16, Decamp16}. We therefore expect very different momentum distributions and trap-level occupancies in the different phases of the Bose-Fermi chain as will be shown in the following.

Momentum distributions of nine bosons (solid) and eight fermions (dashed) are shown in Fig.~\ref{fig-momentum}. Both momentum distributions resemble Gaussian distributions in the Ising AFM phase, Fig.~\ref{fig-momentum}(a), as expected for a Wigner crystal \cite{Deuretzbacher10}. This is a result of the comparatively large distance between the particles of the same kind; see Fig.~\ref{fig-density}(a). In the Heisenberg AFM phase, Fig.~\ref{fig-momentum}(b), the bosonic and fermionic distributions look like those of the corresponding spin-1/2 particles \cite{Yang15, Deuretzbacher16}. By contrast, in the Heisenberg FM phase, Fig.~\ref{fig-momentum}(e), both distributions are much broader, as expected for the highest excited states of the corresponding spin-1/2 particles \cite{Deuretzbacher16}. Indeed, the FM ground state of the Heisenberg Hamiltonian (\ref{eq-Heisenberg}) for $J_i^{(BF)} < 0$ is the highest excited state for $J_i^{(BF)} > 0$. Moreover, the ground state of Eq.~(\ref{eq-XXZ}) for $J_i^{(BB)} = 10 \, J_i^{(BF)} < 0$, which has the momentum distribution shown in Fig.~\ref{fig-momentum}(a), is the highest excited state for $J_i^{(BB)} = 10 \, J_i^{(BF)} > 0$ and the ground state for $J_i^{(BB)} = J_i^{(BF)} < 0$, which has the momentum distribution shown in Fig.~\ref{fig-momentum}(c), is the highest excited state for $J_i^{(BB)} = J_i^{(BF)} > 0$. This is the reason for the broader momentum distributions in the left column of Fig.~\ref{fig-momentum}. Finally, in the Ising FM phase, one has a Tonks-Girardeau gas in the center and noninteracting fermions at the edges of the trap, Fig.~\ref{fig-density}(f). Therefore, one expects that the momentum distributions of that phase resemble those of a Tonks-Girardeau gas and noninteracting fermions, Fig.~\ref{fig-momentum}(f). The distributions are, however, broader than those of Fig.~\ref{fig-momentum}(b), since the particles are located in a smaller trap volume.

\begin{figure}
\begin{center}
\includegraphics[width = \columnwidth]{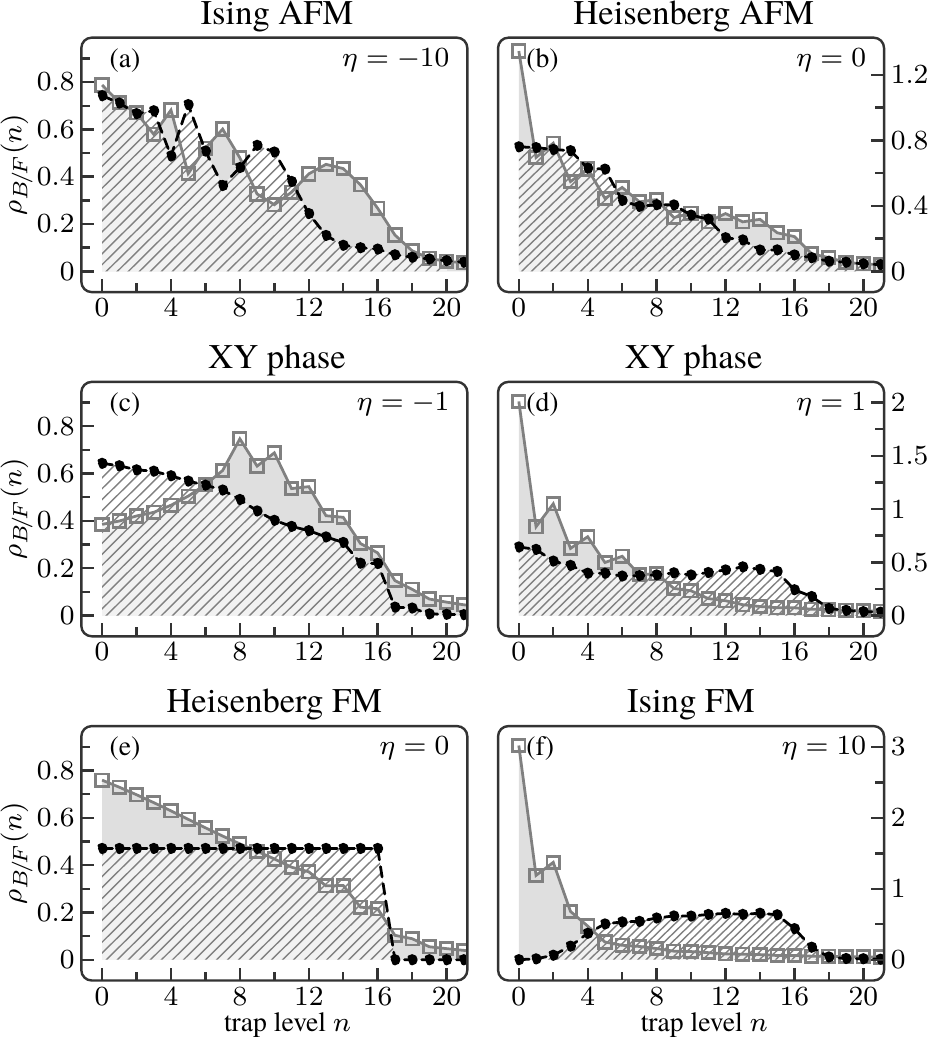}
\caption{Occupancies of the harmonic-trap levels of nine bosons (solid) and eight fermions (dashed) for $g_{BF}<0$ (left) and $g_{BF}>0$ (right). See the text for a description of the different phases.}
\label{fig-occupancy}
\end{center}
\end{figure}

The occupancies of the harmonic-trap levels of nine bosons (solid) and eight fermions (dashed) are shown in Fig.~\ref{fig-occupancy}. In the Ising AFM phase, Fig.~\ref{fig-occupancy}(a), both occupancies oscillate out of phase around the same average broad distribution. The higher oscillator orbitals ($n>10$) are preferably occupied by the bosons, since the bosonic density is broader [Fig.~\ref{fig-density}(a)]. The separation of the bosons from the fermions in the Ising FM phase is also manifest in the occupancies, Fig.~\ref{fig-occupancy}(f). The bosons preferably occupy the lower oscillator orbitals ($n<4$) and the fermions the higher ones ($n>4$). In the Heisenberg AFM phase, Fig.~\ref{fig-occupancy}(b), the bosonic and fermionic occupancies are almost equal. By contrast, in the Heisenberg FM phase, Fig.~\ref{fig-occupancy}(e), the fermions occupy the lowest 17 orbitals with $8/17=0.47$ fermions, whereas the bosonic distribution decreases roughly linearly. Finally, once again, the occupancies in the XY phases, Figs.~\ref{fig-occupancy}(c,d), resemble the behavior of the corresponding densities, Figs.~\ref{fig-density}(c,d).

In closing this section, we note that the variance of the total momentum distribution, $\langle k^2 \rangle = \int dk k^2 (\rho_F(k) + \rho_B(k)) = N^2 / (2 l^2)$, is independent of the pseudospin configuration of the Bose-Fermi chain. Similarly, the total energy of the Bose-Fermi chain, calculated from the total trap-level occupancies, $\sum_{n=0}^\infty (n + 1/2) (\rho_F(n) + \rho_B(n)) = N^2 / 2$, is independent of the configuration of the bosons and fermions. This follows from the fact that all pseudospin configurations have the same total energy at $g_{BB} = g_{BF} = \infty$ and from the virial theorem.

\section{Summary}

We have presented a spin-chain model for Bose-Fermi mixtures with nearly infinite BB and BF $\delta$ interactions. The model is based on a mapping to states of the form $|m_1, \dotsc, m_N\rangle$ with $m_i = B, F$ and to the wave function of spinless noninteracting fermions. We checked the model by comparing with an exact diagonalization of the full few-body Hamiltonian in the strongly interacting regime. Using the spin-chain model, we determined the ground-state phases of the Bose-Fermi mixture and calculated the densities, momentum distributions, and occupancies of the harmonic-trap levels for up to 17 particles. We found, in particular, AFM and FM order and a demixing of the bosons and fermions. However, we found no demixing for equally strong BB and BF interactions in agreement with earlier calculations \cite{Imambekov06}.

\section*{\uppercase{Acknowledgments}}

This work was supported by the DFG [Projects No. SA 1031/7-1, No. RTG 1729, and No. CRC 1227 (DQ-mat), Sub-Project A02], the Cluster of Excellence QUEST, the Swedish Research Council, and NanoLund.

\begin{appendix}

\section{Sector Wave Functions and Permutations}
\label{app-sector-wf}

\subsection{Sector wave functions}

In the regime of infinite BB and BF repulsion, $g_{BB} = g_{BF} = \infty$, the many-body wave function must vanish whenever two particle coordinates are equal, i.e., $\psi(z_1, \dotsc, z_N) = 0$ if $z_i=z_j$. This condition is fulfilled by the wave function of $N$ spinless noninteracting fermions, $\psi_F$, but also by its absolute value $|\psi_F|$, which describes $N$ spinless bosons with infinite $\delta$ repulsion \cite{Girardeau60}. Additionally, we may restrict $|\psi_F|$ to a particular sector of the configuration space $\mathbb R^N$, $z_{P(1)} < \dotsb < z_{P(N)}$, in order to describe spinless distinguishable particles with infinite $\delta$ repulsion and particle ordering $z_{P(1)} < \dotsb < z_{P(N)}$. Here, $P$ denotes an arbitrary permutation of $\underline N = \{1, \dotsc, N\}$. The resulting wave function, denoted by $|P\rangle$, is given by \cite{Deuretzbacher08}
\begin{equation} \label{eq-sector-wf}
\langle z_1, \dotsc , z_N | P \rangle = \sqrt{N!} \, \theta (z_{P(1)}, \dotsc, z_{P(N)}) |\psi_F| .
\end{equation}
Here, $\theta (z_{P(1)}, \dotsc, z_{P(N)}) = 1$ if $z_{P(1)} < \dotsb < z_{P(N)}$ and zero otherwise.

The sector wave functions~(\ref{eq-sector-wf}) are by definition orthonormal, i.e., $\langle P | P' \rangle = \delta_{P,P'}$, and they have further favorable properties. For example, the action of a permutation operator $\hat P$ on a sector wave function $|P'\rangle$ is given by \cite{Deuretzbacher16}
\begin{equation}
\hat P |P'\rangle = |P \circ P'\rangle .
\end{equation}
The permutation operator $\hat P$ of a permutation $P$ acts on a many-body state $|\alpha_1, \dotsc, \alpha_N\rangle$ in the following way:
\begin{equation}
\hat P |\alpha_1\rangle_1 \dotsm |\alpha_N\rangle_N = |\alpha_1\rangle_{P(1)} \dotsm |\alpha_N\rangle_{P(N)} .
\end{equation}
That is, $\hat P$ permutes the particle indices of a many-body state according to the prescription $1 \rightarrow P(1), \dotsc, N \rightarrow P(N)$.

To describe a Bose-Fermi mixture with infinite BB and BF repulsion, ${g_{BB} = g_{BF} = \infty}$, one has to symmetrize the sector wave functions $|P\rangle$ with respect to the bosonic coordinates, $z_1, \dotsc, z_{N_B}$, and to antisymmetrize with respect to the fermionic ones, $z_{N_B+1}, \dotsc, z_N$. We therefore define \cite{Fang11}
\begin{equation} \label{eq-sector-basis}
\hat P | B, \dotsc, B, F, \dotsc, F \rangle \equiv \sqrt{N_B!N_F!} \, S_+ S_- \, | P^{-1} \rangle .
\end{equation}
Here, $S_+ = (1/N_B!) \sum_{P'} \hat P'$ is a symmetrization operator, where the sum runs over all permutations $P'$ of $\underline {N_B} = \{1, \dotsc, N_B\}$, $S_- = (1/N_F!) \sum_{P''} (-1)^{P''} \hat P''$ is an antisymmetrization operator, where the sum runs over all permutations $P''$ of $\underline N - \underline {N_B} = \{N_B+1, \dotsc, N\}$, and $P^{-1}$ is the inverse of the permutation $P$. Furthermore, we specify to use only those initial sector wave functions $|P^{-1}\rangle$, for which the bosonic and fermionic coordinates are each in ascending order. This is necessary, since otherwise two sector wave functions, which differ only by the transposition of two fermionic coordinates, would have a different sign. The requirement is fulfilled if we move the $B$ at position $N_B$ in the initial state $| B, \dotsc, B, F, \dotsc, F \rangle$ to the new position $i_{N_B}$ with $N_B \leq i_{N_B} \leq N$, the $B$ at position $N_B-1$ to the new position $i_{N_B-1}$ with $N_B-1 \leq i_{N_B-1} < i_{N_B}$, and so forth.

\subsection{Permutations}

We use the cycle notation to specify a permutation. For example, the permutation $P_{(\alpha,\beta,\gamma)}$ permutes the numbers $\alpha,\beta,\gamma$ according to the prescription $\alpha \rightarrow \beta \rightarrow \gamma \rightarrow \alpha$. Moreover, we neglect the parentheses if a permutation consists of only one cycle, i.e., $P_{(\alpha,\beta,\gamma)} = P_{\alpha,\beta,\gamma}$. The corresponding unitary operator that permutes the particle indices $\alpha,\beta,\gamma$ of a many-body state according to the same rule is denoted by $\hat P_{\alpha,\beta,\gamma}$. We also note that a cyclic permutation of $\alpha$, $\beta$, and $\gamma$ does not change the cycle, i.e., $P_{\alpha,\beta,\gamma} = P_{\gamma,\alpha,\beta} = P_{\beta,\gamma,\alpha}$.

The cycle $P_{\alpha,\beta,\gamma}$ is the composition of two transpositions $P_{\alpha,\beta}$ and $P_{\beta,\gamma}$, $P_{\alpha,\beta,\gamma} = P_{\alpha,\beta} \circ P_{\beta,\gamma}$. The corresponding cycle operator $\hat P_{\alpha,\beta,\gamma}$ is the product of two transposition operators $\hat P_{\alpha,\beta}$ and $\hat P_{\beta,\gamma}$, $\hat P_{\alpha,\beta,\gamma} = \hat P_{\alpha,\beta} \hat P_{\beta,\gamma}$. The inverse of the cycle operator $\hat P_{\alpha,\beta,\gamma}$ is therefore given by $\hat P_{\alpha,\beta,\gamma}^{-1} = (\hat P_{\alpha,\beta} \hat P_{\beta,\gamma})^{-1} = \hat P_{\gamma,\beta} \hat P_{\beta,\alpha} = \hat P_{\gamma,\beta,\alpha}$, i.e., the particle indices appear in the inverse cycle operator in the inverse order.

The identity permutation is denoted by ``$\mathrm{id}$'' and the corresponding operator by $\openone$. A particular cycle is the loop permutation, which is defined by
\begin{equation}
P_{i,\dotsc,j} = \left\{
\begin{aligned}
& P_{i,i+1,\dotsc,j-1,j} \quad \text{for} \quad i<j \\
& \mathrm{id} \mspace{106.5mu} \text{for} \quad i=j \\
& P_{i,i-1,\dotsc,j+1,j} \quad \text{for} \quad i>j .
\end{aligned}
\right.
\end{equation}
The loop permutation is therefore a composition of transpositions of consecutive integers, $P_{i,\dotsc,j} = P_{i,i+1} \circ P_{i+1,i+2} \circ \, \dotsb \, \circ P_{j-2,j-1} \circ P_{j-1,j}$ (assuming $i<j$) and the loop permutation operator $\hat P_{i,\dotsc,j}$ is a product of transpositions of neighboring particles, $\hat P_{i,\dotsc,j} = \hat P_{i,i+1} \hat P_{i+1,i+2} \dotsm \hat P_{j-2,j-1} \hat P_{j-1,j}$. The loop permutation operator $\hat P_{i,\dotsc,j}$ therefore moves the particle at position~$j$ to position~$i$.

\subsection{Basis of a two-boson two-fermion mixture}

The goal of this section is to clarify definition~(\ref{eq-sector-basis}). A basis of a mixture of two bosons and two fermions (2B2F mixture) is given by
\begin{eqnarray}
& & \mspace{-30mu} |B,B,F,F\rangle , \quad |B,F,B,F\rangle , \quad |B,F,F,B\rangle , \nonumber \\
& & \mspace{-30mu} |F,B,F,B\rangle , \mspace{20mu} |F,F,B,B\rangle , \mspace{20mu} |F,B,B,F\rangle .
\end{eqnarray}
The first basis state is, according to Eq.~(\ref{eq-sector-basis}), constructed by means of the sector wave function $| \mathrm{id} \rangle$ that corresponds to the identity permutation,
\begin{equation}
|B,B,F,F\rangle = \frac{1}{2} (\openone + \hat P_{1,2}) (\openone - \hat P_{3,4}) | \mathrm{id} \rangle .
\end{equation}
The second basis state is obtained from the first one by transposing the second and third particle. Therefore, we obtain
\begin{eqnarray}
|B,F,B,F\rangle & = & \hat P_{2,3} |B,B,F,F\rangle \nonumber \\
& = & \frac{1}{2} (\openone + \hat P_{1,2}) (\openone - \hat P_{3,4}) | P_{2,3} \rangle .
\end{eqnarray}
The third basis state is obtained from the first one by moving the second $B$ to the fourth position. This is achieved by applying the loop permutation operator $\hat P_{4,3,2}$. The inverse of this loop permutation operator is given by $\hat P_{4,3,2}^{-1} = \hat P_{2,3,4}$. We therefore obtain, using Eq.~(\ref{eq-sector-basis}),
\begin{eqnarray}
|B,F,F,B\rangle & = & \hat P_{4,3,2} |B,B,F,F\rangle \nonumber \\
& = & \frac{1}{2} (\openone + \hat P_{1,2}) (\openone - \hat P_{3,4}) | P_{2,3,4} \rangle .
\end{eqnarray}
Note that the bosonic ($z_1, z_2$) and fermionic coordinates ($z_3, z_4$) are each in ascending order in the initial sector wave function $| P_{2,3,4} \rangle$, since $z_1<z_3<z_4<z_2$, as required.

The fourth basis state is obtained from the first one by moving the second $B$ to the fourth position and then the first $B$ to the second position, i.e., by applying $\hat P_{1,2} \hat P_{4,3,2} = \hat P_{1,2,4,3}$. Using $\hat P_{1,2,4,3}^{-1} = \hat P_{3,4,2,1} = \hat P_{2,1,3,4}$ we obtain
\begin{equation}
|F,B,F,B\rangle = \frac{1}{2} (\openone + \hat P_{1,2}) (\openone - \hat P_{3,4}) | P_{2,1,3,4} \rangle .
\end{equation}
The fifth basis state is obtained from the first one by moving the second $B$ to the fourth position and then the first $B$ to the third position, i.e., by applying $\hat P_{3,2,1} \hat P_{4,3,2} = \hat P_{(1,3)(2,4)}$. Using $\hat P_{(1,3)(2,4)}^{-1} = \hat P_{(1,3)(2,4)}$ we obtain
\begin{equation}
|F,F,B,B\rangle = \frac{1}{2} (\openone + \hat P_{1,2}) (\openone - \hat P_{3,4}) | P_{(1,3)(2,4)} \rangle .
\end{equation}
The sixth basis state is finally obtained from the first one by moving the second $B$ to the third position and then the first $B$ to the second position, i.e., by applying $\hat P_{2,1} \hat P_{3,2} = \hat P_{1,2,3}$. Using $\hat P_{1,2,3}^{-1} = \hat P_{3,2,1}$ we obtain
\begin{equation}
|F,B,B,F\rangle = \frac{1}{2} (\openone + \hat P_{1,2}) (\openone - \hat P_{3,4}) | P_{3,2,1} \rangle .
\end{equation}

\section{Effective Hamiltonian}
\label{app-Heff}

Here, we perform a perturbative calculation of a strongly interacting 2B1F mixture up to linear order in $1/g_{BB}$ and $1/g_{BF}$. The matrix elements of the Hamiltonian in the degenerate ground-state manifold are shown to agree with Eq.~(\ref{eq-Heff}). The basis states of the 2B1F mixture are [see Eq.~(\ref{eq-sector-basis})]
\begin{eqnarray}
|1\rangle & \!\! := |B,B,F\rangle & \!\! = \frac{1}{\sqrt{2}} (\openone + \hat P_{1,2}) |\mathrm{id}\rangle , \\
|2\rangle & \!\! := |B,F,B\rangle & \!\! = \frac{1}{\sqrt{2}} (\openone + \hat P_{1,2}) |P_{2,3}\rangle , \\
|3\rangle & \!\! := |F,B,B\rangle & \!\! = \frac{1}{\sqrt{2}} (\openone + \hat P_{1,2}) |P_{3,2,1}\rangle .
\end{eqnarray}
The Hamiltonian of the 2B1F mixture is given by [see Eq.~(\ref{eq-HBF})]
\begin{eqnarray}
H & = & \sum_{i=1}^3 \left[ -\frac{\hbar^2}{2M} \frac{\partial^2}{\partial z_i^2} + V(z_i) \right] + g_{BB} \delta(z_1-z_2) \nonumber \\
& & + \: g_{BF} \delta (z_1 - z_3) + g_{BF} \delta (z_2 - z_3) .
\end{eqnarray}
The matrix element $\langle 1|H|1\rangle$ is therefore given by
\begin{equation}
\langle 1|H|1\rangle = \frac{1}{2} \langle \mathrm{id}| (\openone + \hat P_{1,2}) H (\openone + \hat P_{1,2}) |\mathrm{id}\rangle .
\end{equation}
$H$ is symmetric under the exchange of the first and second particle. $H$ therefore commutes with $(\openone + \hat P_{1,2})$. Moreover, $(\openone + \hat P_{1,2}) (\openone + \hat P_{1,2}) = 2 (\openone + \hat P_{1,2})$ and therefore
\begin{equation}
\langle 1|H|1\rangle = \langle \mathrm{id}| H |\mathrm{id}\rangle + \langle \mathrm{id}| H |P_{1,2}\rangle .
\end{equation}
Let us calculate an arbitrary matrix element $\langle P|H|P'\rangle$ in the vicinity of $(1/g_{BB},1/g_{BF}) = (0,0)$. Performing a Taylor expansion up to first order in $1/g_{BB}$ and $1/g_{BF}$, we obtain~\cite{Deuretzbacher14}
\begin{eqnarray}
& & \mspace{-50mu} \langle P|H|P'\rangle = E_F \delta_{P,P'} \nonumber \\
& & \mspace{-30mu} - \frac{1}{g_{BB}} \lim_{g_{BB} \rightarrow +\infty} \left( g_{BB}^2 \langle P^{(g_{BB})}| \frac{dH}{dg_{BB}} |P'^{(g_{BB})}\rangle \right) \nonumber \\
& & \mspace{-30mu} - \frac{1}{g_{BF}} \lim_{g_{BF} \rightarrow +\infty} \left( g_{BF}^2 \langle P^{(g_{BF})}| \frac{dH}{dg_{BF}} |P'^{(g_{BF})}\rangle \right)
\end{eqnarray}
with $\frac{dH}{dg_{BB}} = \delta(z_1-z_2)$ and $\frac{dH}{dg_{BF}} = \delta (z_1 - z_3) + \delta (z_2 - z_3)$. $|P^{(g)}\rangle$ is the ground state of $N$ spinless bosons with strong $\delta$ repulsion restricted to the sector $z_{P(1)} < \dotsb < z_{P(N)}$ \cite{Deuretzbacher14}. Furthermore, using the boundary condition
\begin{eqnarray}
& & \mspace{-30mu} \left( \frac{\partial}{\partial z_i} - \frac{\partial}{\partial z_j} \right) \psi \big|_{z_i=z_j+} - \left( \frac{\partial}{\partial z_i} - \frac{\partial}{\partial z_j} \right) \psi \big|_{z_i=z_j-} \nonumber \\
& & \mspace{-10mu} = \frac{2Mg}{\hbar^2} \psi \big|_{z_i=z_j} ,
\end{eqnarray}
one finds
\begin{equation}
\lim_{g \rightarrow +\infty} \left( g^2 \langle P^{(g)}| \delta(z_i-z_j) |P'^{(g)}\rangle \right) = C_{i,j}^{P,P'}
\end{equation}
with
\begin{eqnarray}
\mspace{-50mu} C_{i,j}^{P,P'} & \!\! = & \! \frac{N!\hbar^4}{M^2} \int dz_1 \dotsi dz_N \, \delta(z_i-z_j) \left| \frac{\partial \psi_F}{\partial z_i} \right|^2 \nonumber \\
& & \! \times \theta (z_{P(1)},\dotsc,z_{P(N)}) \theta (z_{P'(1)},\dotsc,z_{P'(N)}) .
\end{eqnarray}
As a result, we obtain
\begin{equation}
\langle P|H|P'\rangle = E_F \delta_{P,P'} - \frac{1}{g_{BB}} C_{1,2}^{P,P'} \!\! - \frac{1}{g_{BF}} \! \left( \! C_{1,3}^{P,P'} \!\! + C_{2,3}^{P,P'} \right) \! .
\end{equation}
Applying this to the matrix element $\langle 1|H|1\rangle$, we get
\begin{equation}
\langle 1|H|1\rangle = E_F - 2 J_1^{(BB)} - J_2^{(BF)} ,
\end{equation}
since $C_{1,2}^{\mathrm{id},\mathrm{id}} = C_{1,2}^{\mathrm{id},P_{1,2}} = C_1$, $C_{2,3}^{\mathrm{id},\mathrm{id}} = C_2$, and $C_{1,3}^{\mathrm{id},\mathrm{id}} = C_{1,3}^{\mathrm{id},P_{1,2}} = C_{2,3}^{\mathrm{id},P_{1,2}} = 0$. In a similar way we obtain
\begin{equation}
\langle 1|H|2\rangle = \langle \mathrm{id}| H |P_{2,3}\rangle + \langle \mathrm{id}| H |P_{1,2,3}\rangle = -J_2^{(BF)} ,
\end{equation}
since only $C_{2,3}^{\mathrm{id},P_{2,3}} = C_2$ is nonzero. For the next matrix element, we get
\begin{equation}
\langle 1|H|3\rangle = \langle \mathrm{id}| H |P_{1,3}\rangle + \langle \mathrm{id}| H |P_{3,2,1}\rangle = 0 ,
\end{equation}
since $C_{i,j}^{\mathrm{id},P_{1,3}} = C_{i,j}^{\mathrm{id},P_{3,2,1}} = 0$ for all $1 \leq i < j \leq 3$. The next matrix element becomes
\begin{eqnarray}
\langle 2|H|2\rangle & = & \langle P_{2,3}| H |P_{2,3}\rangle + \langle P_{2,3}| H |P_{1,2,3}\rangle \nonumber \\
& = & E_F - J_1^{(BF)} - J_2^{(BF)} ,
\end{eqnarray}
since only $C_{1,3}^{P_{2,3},P_{2,3}} = C_1$ and $C_{2,3}^{P_{2,3},P_{2,3}} = C_2$ are nonzero. Finally we obtain
\begin{equation}
\langle 2|H|3\rangle = \langle P_{2,3}| H |P_{1,3}\rangle + \langle P_{2,3}| H |P_{3,2,1}\rangle = -J_1^{(BF)}
\end{equation}
and
\begin{eqnarray}
\langle 3|H|3\rangle & = & \langle P_{3,2,1}| H |P_{1,3}\rangle + \langle P_{3,2,1}| H |P_{3,2,1}\rangle \nonumber \\
& = & E_F - 2 J_2^{(BB)} - J_1^{(BF)} ,
\end{eqnarray}
since only $C_{1,3}^{P_{2,3},P_{3,2,1}} = C_{1,3}^{P_{3,2,1},P_{3,2,1}} = C_1$ and $C_{1,2}^{P_{3,2,1},P_{1,3}} = C_{1,2}^{P_{3,2,1},P_{3,2,1}} = C_2$ are nonzero. The same matrix elements are obtained using $H_\mathrm{eff}$, given by Eq.~(\ref{eq-Heff}).

\section{One-Body Density Matrix}
\label{app-one-body}

Here, we calculate the matrix elements of the bosonic and fermionic one-body density matrices of a 1B2F mixture. The basis states of the 1B2F mixture are [see Eq.~(\ref{eq-sector-basis})]
\begin{eqnarray}
|1\rangle & \!\! := |B,F,F\rangle & \!\! = \frac{1}{\sqrt{2}} (\openone - \hat P_{2,3}) |\mathrm{id}\rangle , \\
|2\rangle & \!\! := |F,B,F\rangle & \!\! = \frac{1}{\sqrt{2}} (\openone - \hat P_{2,3}) |P_{1,2}\rangle , \\
|3\rangle & \!\! := |F,F,B\rangle & \!\! = \frac{1}{\sqrt{2}} (\openone - \hat P_{2,3}) |P_{1,2,3}\rangle .
\end{eqnarray}
The bosonic and fermionic one-body density matrix operators read
\begin{equation}
\hat \rho_B(z,z') = |z\rangle_1 \langle z'|_1
\end{equation}
and
\begin{equation}
\hat \rho_F(z,z') = |z\rangle_2 \langle z'|_2 + |z\rangle_3 \langle z'|_3 .
\end{equation}
First, we calculate the matrix elements of the bosonic distribution $\hat \rho_B(z,z')$. One finds
\begin{equation}
\langle 1| \hat \rho_B(z,z') |1\rangle = \langle\mathrm{id}|z\rangle_1 \langle z'|_1 (|\mathrm{id}\rangle - |P_{2,3}\rangle) .
\end{equation}
Only those matrix elements of the form $\langle\mathrm{id}|z\rangle_i \langle z'|_i |P\rangle$ are nonzero for which $P=P_{i,\dotsc,j}$. We define
\begin{equation}
\rho^{(i,j)}(z,z') = \langle\mathrm{id}|z\rangle_i \langle z'|_i |P_{i,\dotsc,j}\rangle .
\end{equation}
Using this, we find
\begin{eqnarray}
& & \mspace{-40mu} \langle 1| \hat \rho_B(z,z') |1\rangle = \rho^{(1,1)}(z,z') \nonumber \\
& & \mspace{-20mu} = \langle B,F,F | \rho^{(1,1)}(z,z') |B\rangle_1 \langle B|_1 | B,F,F \rangle .
\end{eqnarray}
The next two matrix elements are given by
\begin{eqnarray}
& & \mspace{-50mu} \langle 1| \hat \rho_B(z,z') |2\rangle = \langle\mathrm{id}|z\rangle_1 \langle z'|_1 (|P_{1,2}\rangle - |P_{3,2,1}\rangle) \nonumber \\
& & \mspace{-30mu} = \rho^{(1,2)}(z,z') \nonumber \\
& & \mspace{-30mu} = \langle B,F,F | \rho^{(1,2)}(z,z') |B\rangle_1 \langle B|_1 \hat P_{1,2} | F,B,F \rangle
\end{eqnarray}
and
\begin{eqnarray}
& & \mspace{-60mu} \langle 1| \hat \rho_B(z,z') |3\rangle = \langle\mathrm{id}|z\rangle_1 \langle z'|_1 (|P_{1,2,3}\rangle - |P_{1,3}\rangle) \nonumber \\
& & \mspace{-40mu} = \rho^{(1,3)}(z,z') \nonumber \\
& & \mspace{-40mu} = \langle B,F,F | \rho^{(1,3)}(z,z') |B\rangle_1 \langle B|_1 \hat P_{1,2,3} | F,F,B \rangle .
\end{eqnarray}
In the next case, we find
\begin{equation}
\langle 2| \hat \rho_B(z,z') |2\rangle = \langle P_{1,2}|z\rangle_1 \langle z'|_1 (|P_{1,2}\rangle - |P_{3,2,1}\rangle) .
\end{equation}
It is easy to show that
\begin{equation} \label{eq-useful-formula}
\langle P | z \rangle_i \langle z' |_i | P' \rangle = \langle \mathrm{id} | z \rangle_{P^{-1}(i)} \langle z' |_{P^{-1}(i)} | P^{-1} \circ P' \rangle .
\end{equation}
Using this, we find
\begin{eqnarray}
& & \mspace{-22mu} \langle 2| \hat \rho_B(z,z') |2\rangle = \langle\mathrm{id}|z\rangle_2 \langle z'|_2 (|\mathrm{id}\rangle - |P_{1,3}\rangle) = \rho^{(2,2)}(z,z') \nonumber \\
& & \mspace{-2mu} = \langle F,B,F | \rho^{(2,2)}(z,z') |B\rangle_2 \langle B|_2 | F,B,F \rangle .
\end{eqnarray}
In the next case, we obtain
\begin{eqnarray}
& & \mspace{-50mu} \langle 2| \hat \rho_B(z,z') |3\rangle = \langle P_{1,2}|z\rangle_1 \langle z'|_1 (|P_{1,2,3}\rangle - |P_{1,3}\rangle) \nonumber \\
& & \mspace{-30mu} = \langle\mathrm{id}|z\rangle_2 \langle z'|_2 (|P_{2,3}\rangle - |P_{3,2,1}\rangle) = \rho^{(2,3)}(z,z') \nonumber \\
& & \mspace{-30mu} = \langle F,B,F | \rho^{(2,3)}(z,z') |B\rangle_2 \langle B|_2 \hat P_{2,3} | F,F,B \rangle .
\end{eqnarray}
The last matrix element is given by
\begin{eqnarray}
& & \mspace{-40mu} \langle 3| \hat \rho_B(z,z') |3\rangle = \langle P_{1,2,3}|z\rangle_1 \langle z'|_1 (|P_{1,2,3}\rangle - |P_{1,3}\rangle) \nonumber \\
& & \mspace{-20mu} = \langle\mathrm{id}|z\rangle_3 \langle z'|_3 (|\mathrm{id}\rangle - |P_{1,2}\rangle) = \rho^{(3,3)}(z,z') \nonumber \\
& & \mspace{-20mu} = \langle F,F,B | \rho^{(3,3)}(z,z') |B\rangle_3 \langle B|_3 | F,F,B \rangle .
\end{eqnarray}
One sees that the matrix elements of the bosonic one-body density matrix agree with those of Eqs.~(\ref{eq-one-body})--(\ref{eq-rhoij}). Now, we calculate the matrix elements of the fermionic distribution $\hat \rho_F(z,z')$. The first two matrix elements read
\begin{widetext}
\begin{eqnarray}
\mspace{-10mu} \langle 1| \hat \rho_F(z,z') |1\rangle & \!\! = & \!\! \langle\mathrm{id}| (|z\rangle_2 \langle z'|_2 + |z\rangle_3 \langle z'|_3) (|\mathrm{id}\rangle - |P_{2,3}\rangle) = \rho^{(2,2)}(z,z') - \rho^{(2,3)}(z,z') + \rho^{(3,3)}(z,z') - \rho^{(3,2)}(z,z') \nonumber \\
& \!\! = & \!\! \langle B,F,F | \Bigl( \rho^{(2,2)}(z,z') |F\rangle_2 \langle F|_2 - \rho^{(2,3)}(z,z') |F\rangle_2 \langle F|_2 \hat P_{2,3} \nonumber \\
& & \mspace{73mu} + \rho^{(3,3)}(z,z') |F\rangle_3 \langle F|_3 - \rho^{(3,2)}(z,z') |F\rangle_3 \langle F|_3 \hat P_{3,2} \Bigr) | B,F,F \rangle
\end{eqnarray}
and
\begin{eqnarray}
\langle 1| \hat \rho_F(z,z') |2\rangle & \!\! = & \!\! \langle\mathrm{id}| (|z\rangle_2 \langle z'|_2 + |z\rangle_3 \langle z'|_3) (|P_{1,2}\rangle - |P_{3,2,1}\rangle) = \rho^{(2,1)}(z,z') - \rho^{(3,1)}(z,z') \nonumber \\
& \!\! = & \!\! \langle B,F,F | \left( \rho^{(2,1)}(z,z') |F\rangle_2 \langle F|_2 \hat P_{2,1} - \rho^{(3,1)}(z,z') |F\rangle_3 \langle F|_3 \hat P_{3,2,1} \right) | F,B,F \rangle .
\end{eqnarray}
The next matrix element is zero,
\begin{equation}
\langle 1| \hat \rho_F(z,z') |3\rangle = \langle\mathrm{id}| (|z\rangle_2 \langle z'|_2 + |z\rangle_3 \langle z'|_3) (|P_{1,2,3}\rangle - |P_{1,3}\rangle) = 0 .
\end{equation}
Using Eq.~(\ref{eq-useful-formula}), we obtain for the last three matrix elements
\begin{eqnarray}
\mspace{-44mu} \langle 2| \hat \rho_F(z,z') |2\rangle & \!\! = & \!\! \langle P_{1,2}| (|z\rangle_2 \langle z'|_2 + |z\rangle_3 \langle z'|_3) (|P_{1,2}\rangle - |P_{3,2,1}\rangle) = \langle\mathrm{id}| (|z\rangle_1 \langle z'|_1 + |z\rangle_3 \langle z'|_3) (|\mathrm{id}\rangle - |P_{1,3}\rangle) \nonumber \\
& \!\! = & \!\! \rho^{(1,1)}(z,z') + \rho^{(3,3)}(z,z') = \langle F,B,F | \left( \rho^{(1,1)}(z,z') |F\rangle_1 \langle F|_1 + \rho^{(3,3)}(z,z') |F\rangle_3 \langle F|_3 \right) | F,B,F \rangle ,
\end{eqnarray}
\begin{eqnarray}
\langle 2| \hat \rho_F(z,z') |3\rangle & \!\! = & \!\! \langle P_{1,2}| (|z\rangle_2 \langle z'|_2 + |z\rangle_3 \langle z'|_3) (|P_{1,2,3}\rangle - |P_{1,3}\rangle) = \langle\mathrm{id}| (|z\rangle_1 \langle z'|_1 + |z\rangle_3 \langle z'|_3) (|P_{2,3}\rangle - |P_{3,2,1}\rangle) \nonumber \\
& \!\! = & \!\! \rho^{(3,2)}(z,z') - \rho^{(3,1)}(z,z') = \langle F,B,F | \! \left( \rho^{(3,2)}(z,z') |F\rangle_3 \langle F|_3 \hat P_{3,2} - \rho^{(3,1)}(z,z') |F\rangle_3 \langle F|_3 \hat P_{3,2,1} \! \right) \! | F,F,B \rangle , \nonumber \\
& &
\end{eqnarray}
and
\begin{eqnarray}
\mspace{-10mu} \langle 3| \hat \rho_F(z,z') |3\rangle & \!\! = & \!\! \langle P_{1,2,3}| (|z\rangle_2 \langle z'|_2 + |z\rangle_3 \langle z'|_3) (|P_{1,2,3}\rangle - |P_{1,3}\rangle) = \langle\mathrm{id}| (|z\rangle_1 \langle z'|_1 + |z\rangle_2 \langle z'|_2) (|\mathrm{id}\rangle - |P_{1,2}\rangle) \nonumber \\
& \!\! = & \!\! \rho^{(1,1)}(z,z') - \rho^{(1,2)}(z,z') + \rho^{(2,2)}(z,z') - \rho^{(2,1)}(z,z') \nonumber \\
& \!\! = & \!\! \langle F,F,B | \Bigl( \rho^{(1,1)}(z,z') |F\rangle_1 \langle F|_1 - \rho^{(1,2)}(z,z') |F\rangle_1 \langle F|_1 \hat P_{1,2} \nonumber \\
& & \mspace{70mu} + \rho^{(2,2)}(z,z') |F\rangle_2 \langle F|_2 - \rho^{(2,1)}(z,z') |F\rangle_2 \langle F|_2 \hat P_{2,1} \Bigr) | F,F,B \rangle .
\end{eqnarray}
Again, one sees that the matrix elements of the fermionic one-body density matrix agree with those of Eqs.~(\ref{eq-one-body})--(\ref{eq-rhoij}).
\end{widetext}

\end{appendix}

\bibliographystyle{prsty}

\begin{thebibliography}{}
\bibitem{Comment1} In a quasi-1D trap with a tight transverse harmonic confinement with frequency $\omega_\perp$, the inverse strength, $1/g$, of the 1D $\delta$ interaction fulfills $2 \hbar \omega_\perp l_\perp / g = l_\perp / a_{3D} - C$, with $l_\perp = \sqrt{\hbar / (m \omega_\perp)}$, $a_{3D}$ the bare 3D scattering length, and $C \simeq 1.0326 \dotsc$~\cite{Olshanii98}. Hence, by changing either $a_{3D}$ or the transversal confinement, $\hbar \omega l / g$ may be adiabatically changed from an initial small positive (Tonks regime) to a small negative value (super-Tonks regime). The super-Tonks state is a highly excited state but remains metastable due to the small overlap with the strongly bound states. Around the confinement-induced resonance, the charge (i.e. density) degrees of freedom are frozen and well approximated by the wave function of spinless noninteracting fermions, while the spin degrees of freedom are described by the spin-chain model.
\end{thebibliography}

\end{document}